\documentstyle[prb,preprint,aps]{revtex}
\draft
\tighten
\newcommand{\ro}{\bbox{\rho}}
\newcommand{\bpi}{\bbox{\pi}}
\newcommand{\bPi}{\bbox{\Pi}}
\newcommand{\bn}{\bbox{\nabla}}
\begin{document}
\title{Charged two-dimensional magnetoexciton and
             two-mode squeezed vacuum states}
\author{A. B. Dzyubenko\cite{ABD}}
\address{Department of Physics, University at Buffalo, SUNY,
          Buffalo, NY 14260, USA}
\date{JETP Lett. {\bf 64}, 318 (2001)}
\maketitle
\begin{abstract}
A novel unitary transformation of the Hamiltonian that allows
one to partially separate the center-of-mass motion for charged
electron-hole systems in a magnetic field is presented.
The two-mode squeezed oscillator states that appear at the
intermediate stage of the transformation are used for constructing
a trial wave function of a two-dimensional (2D) charged magnetoexciton.
\end{abstract}
\pacs{73.20.Mf,71.35.Ji,73.43.Lp}

A problem of center-of-mass (CM) separation for a quantum-mechanical
system of charged interacting particles in a magnetic field $B$
has been studied by many authors. \cite{G&D,Simon,Bychkov,Hirsch,SSC}
When a charge-to-mass ratio is the same for all particles,
the CM and internal motions decouple \cite{Simon,Bychkov,Hirsch} in $B$.
For a neutral system, the CM coordinates can be separated \cite{G&D,Simon}
in the Schr\"odinger equation. This is associated with the fact that
translations commute for a neutral system in $B$.
In general, only a partial separation of the CM in magnetic fields
is possible. \cite{Simon,Hirsch,SSC}
In this Letter we propose a novel operator approach for
performing such a separation in charged electron-hole ($e$--$h$)
systems in $B$.
This approach can be useful for studying in strong magnetic fields,
e.g., atomic ions with not too large mass ratios \cite{Hirsch}
and charged excitations in 2D electron systems, in particular,
in the fractional quantum Hall effect regime
in the planar geometry. \cite{Ezawa,Pasquier}
In this work, we study in 2D a three-particle problem of two electrons
and one hole in a strong magnetic field, i.e., a negatively charged
magnetoexciton $X^-$
(see Refs.~\onlinecite{SSC,X-th,Dz&S_PRL} and references therein).
We consider an approximate $X^-$ ground state in the form, which is related
to the two-mode squeezed \cite{Klauder} oscillator vacuum states.

The Hamiltonian describing the 2D three-particle $2e$--$h$ complex
in a perpendicular magnetic field ${\bf B}$ is
$H = H_0 + H_{\rm int}$, where the free-particle part is given by
\begin{equation}
                \label{H_0} 
  H_0 = \sum_{i=1,2} \frac{\hat{\bPi}_{ei}^2}{2m_e} +
                      \frac{\hat{\bPi}_{h}^2 }{2m_h}
    \equiv   \sum_{i=1,2} H_{0e}({\bf r}_i) + H_{0h}({\bf r}_h)  \, ,
\end{equation}
and $\hat{\bPi}_j = -i\hbar \bn_j -  \frac{e_j}{c} {\bf A}({\bf r}_j)$
are kinematic momentum operators.
The interaction Hamiltonian $H_{\rm int} = H_{\rm ee} + H_{\rm eh}$ is
\begin{equation}
            \label{H_int} 
 H_{\rm ee} = \frac{e^2}{\epsilon |{\bf r}_1-{\bf r}_2|} \quad ,  \quad
 H_{\rm eh} = - \sum_{i=1,2} \frac{e^2}{\epsilon |{\bf r}_i-{\bf r}_h|} \, .
\end{equation}
The Hamiltonian $H$ commutes \cite{Simon,Hirsch,Dz&S_PRL}
with the operator of magnetic translations (MT)
$\hat{\bf K} = \sum_{j} \hat{\bf K}_j$,
where
$\hat{\bf K}_j = \hat{\bPi}_j - \frac{e_j}{c} {\bf r}_j \times {\bf B}$.
In the symmetric gauge, ${\bf A} = \frac12 {\bf B} \times {\bf r}$,
the operators satisfy the relation
$\hat{\bf K}_j({\bf B}) = \hat{\bbox \Pi}_j(-{\bf B})$;
independent of the gauge, $\hat{\bf K}_j$ and $\hat{\bbox \Pi}_j$ commute.
The important feature of  $\hat{\bf K}$
and $\hat{\bPi} = \sum_{j} \hat{\bPi}_j$
is the non-commutativity of the components in $B$:
$[\hat{K}_x, \hat{K}_y] = - [\hat{\Pi}_x, \hat{\Pi}_y] =
- i \frac{\hbar B}{c} Q$, where $Q = \sum_j e_j$ is the total charge.
This allows one to introduce the raising and lowering
Bose ladder operators for the whole system \cite{Simon,Hirsch,Dz&S_PRL}
\begin{eqnarray}
  \label{kmin}
  \hat{k}_{\pm} &=& \pm \frac{i}{\sqrt{2}} (\hat{k}_x  \pm i \hat{k}_y)
            \quad , \quad
    [\hat{k}_{+}, \hat{k}_{-}]=-\frac{Q}{|Q|} \, ,   \\
  \label{pimin}
  \hat{\pi}_{\pm} &=& \mp \frac{i}{\sqrt{2}} (\hat{\pi}_x  \pm i \hat{\pi}_y)
            \quad , \quad
    [\hat{\pi}_{+}, \hat{\pi}_{-}]= \frac{Q}{|Q|} \, ,
\end{eqnarray}
where $\hat{{\bf k}} = \sqrt{c/\hbar B |Q|} \, \hat{\bf K}$,
$\hat{{\bpi}} =  \sqrt{c/\hbar B |Q|} \, \hat{\bPi}$,
and the phases of the operators (\ref{kmin}) and (\ref{pimin})
can be chosen arbitrary.
The operator $\hat{{\bf k}}^2$ has the discrete oscillator
eigenvalues $2k+1$, $k=0, 1, \ldots$ that are associated \cite{Simon,Hirsch}
with the guiding center of a charged complex in $B$.
The values of $k$ can be used, \cite{remark}
together with the total angular
momentum projection $M_z$ and the electron, $S_e$, and hole, $S_h$,
spin quantum numbers, for the classification of states;  \cite{Dz&S_PRL}
the exact eigenenergies are degenerate \cite{Simon,Hirsch} in $k$.

In terms of the {\em single-particle\/} Bose ladder intra Landau level (LL)
operators \cite{SSC,Ezawa}
$B^{\dag}_{e}({\bf r}_j) =
 -i \sqrt{c/2 \hbar B e} \, (\hat{ K }_{jx} - i \hat{ K }_{jy})$
for the electrons and
$B^{\dag}_{h}({\bf r}_h) =
 - i \sqrt{c/2 \hbar B e} \, (\hat{ K }_{hx} + i \hat{ K }_{hy})$
for the hole, the raising operator takes the form
$\hat{k}_{-} =
B_e^{\dag}({\bf r}_1) + B_e^{\dag}({\bf r}_2) - B_h({\bf r}_h)$.
One needs to diagonalize $\hat{k}_{-}$ in order to maintain
the exact MT symmetry. This can be achieved by performing
first an orthogonal transformation \cite{Bychkov,Hirsch}
of the electron coordinates
$\{ {\bf r}_1,{\bf r}_2,{\bf r}_{h}\} \rightarrow
 \{ {\bf r},{\bf R}, {\bf r}_h \}$,
where ${\bf r} = ({\bf r}_1 - {\bf r}_2)/\sqrt{2}$, and
${\bf R} = ({\bf r}_1 + {\bf r}_2)/\sqrt{2}$
are the electron relative and CM coordinates.
In these coordinates
$\hat{k}_{-} = \sqrt{2}\,B_e^{\dag}({\bf R}) - B_h({\bf r}_h)$
and can be considered to be a new Bose ladder operator
generated by the Bogoliubov transformation \cite{SSC}
\begin{equation}
        \label{Bog1}
 \tilde{B}_e^{\dag}({\bf R}) \equiv
   u B^{\dag}_{e}({\bf R}) - v B_{h}({\bf r}_h)
 = \tilde{S} B_e^{\dag}({\bf R}) \tilde{S}^{\dag} \, ,
\end{equation}
where the unitary operator \cite{SSC,Ezawa,Klauder}
$\tilde{S}  =   \exp ( \Theta \tilde{\cal L} )$
and the generator
$\tilde{\cal L}  =  B^{\dag}_{h}({\bf r}_h) B^{\dag}_{e}({\bf R})
- B_e({\bf R})B_h({\bf r}_h)$.
Here $\Theta$ is the transformation angle and
$u= \cosh \Theta=\sqrt{2}$, $v=\sinh\Theta =1$.
Now we have $\hat{k}_- = \tilde{B}_e^{\dag}$ and
$\hat{{\bf k}}^2 = 2\tilde{B}_e^{\dag} \tilde{B}_e  + 1$.
The second linearly independent creation operator is
\begin{equation}
        \label{Bog2}
 \tilde{B}_h^{\dag}({\bf r}_h) =
            \tilde{S} B^{\dag}_{h}({\bf r}_h) \tilde{S}^{\dag} =
                      u B^{\dag}_{h}({\bf r}_h) - v B_{e}({\bf R})  \, .
\end{equation}
A complete orthonormal basis compatible with both axial and translational
symmetries can be constructed \cite{SSC} as:
\begin{eqnarray}
          \nonumber
  &&     \frac{A_e^{\dag}({\bf r})^{n_r}
               A_e^{\dag}({\bf R})^{n_R}
               A_h^{\dag}({\bf r}_h)^{n_h}
      \tilde{B}_e^{\dag}({\bf R})^k
      \tilde{B}_h^{\dag}({\bf r}_h)^l
             B_e^{\dag}({\bf r})^m |\tilde{0} \rangle}
                {\left(n_r!n_R!n_h!k!l!m! \right)^{1/2}} \\
        \label{basis}
  & & \quad \quad \quad \equiv |n_r n_R n_h ; \widetilde{ k l m} \rangle  \, .
\end{eqnarray}
Here the inter-LL Bose ladder operators are given by
$A^{\dag}_{e}({\bf r}_j) =
 - i \sqrt{c/2 \hbar B e} \, (\hat{ \Pi }_{jx} + i \hat{ \Pi }_{jy})$
and
$A^{\dag}_{h}({\bf r}_h) =
  - i \sqrt{c/2 \hbar B e} \, (\hat{ \Pi }_{hx} - i \hat{ \Pi }_{hy})$;
the explicit form is given in, e.g., Refs.~\onlinecite{SSC,Ezawa}.
The tilde sign shows that the transformed
vacuum state $|\tilde{0}\rangle$ (see below) and the transformed
operators (\ref{Bog1}) and (\ref{Bog2}) are involved.
In (\ref{basis}) the oscillator quantum number is fixed and equals $k$,
while $M_z= n_r + n_R- n_h -k + l - m$.
The permutational symmetry requires that $n_r-m$ should be even (odd)
for electron spin-singlet $S_e=0$ (triplet $S_e=1$ states);
see Ref.~\onlinecite{SSC} for more details.

The transformation introduces a new vacuum state
$|\tilde{0} \rangle = \tilde{S} |0 \rangle$, for which,
using the normal-ordered form \cite{SSC,Ezawa,Klauder} of $\tilde{S}$,
one obtains
\begin{equation}
         \label{vacuum}
 |\tilde{0} \rangle = \tilde{S} |0\rangle = \frac{1}{\cosh  \Theta }
  \exp\left[ \tanh \Theta \,
  B^{\dag}_{h}({\bf r}_h) B^{\dag}_{e}({\bf R}) \right] |0\rangle \, .
\end{equation}
The coordinate representation has the form
\begin{equation}
         \label{rvacuum}
   \langle {\bf r} {\bf R} {\bf r}_h | \tilde{0} \rangle =
    \frac{1}{\sqrt{2}\,(2\pi l_B^2)^{3/2}}  \,
   \label{vac-r}
    \exp \left( -\frac{{\bf r}^2 + {\bf R}^2 + {\bf r}_h^2 -
      \sqrt{2}Z^{\ast} z_h }{4 l_B^2} \right) \, ,
\end{equation}
where $l_B=(\hbar c/eB)^{1/2}$ is the magnetic length,
$Z^{\ast} = X-iY$, and $z_h = x_h + iy_h$.
Equation (\ref{rvacuum}) shows that $| \tilde{0} \rangle$ contains
a {\em coherent superposition\/} of an infinite number of $e$- and $h$-
states in zero LL's. In the terminology of quantum optics, \cite{Klauder}
$| \tilde{0} \rangle $ is a {\em two-mode squeezed state};
for particles in a magnetic field the squeezing has a
direct geometrical meaning. \cite{oneparticle}
Indeed, the probability distribution function takes the factored form
\begin{eqnarray}
     \label{vac-prob}
  |\langle {\bf r} {\bf R} {\bf r}_h | \tilde{0} \rangle|^2 =
    \frac{1}{2\pi l_B^2} \exp \left( -\frac{{\bf r}^2}{2l_B^2} \right) \,  &&
           \frac{2+\sqrt{2}}{4 \pi l_B^2}
           \exp \left[ - \frac{2+\sqrt{2}}{8 l_B^2}
           \left(  {\bf R}-{\bf r}_h \right)^2  \right]   \\
               \nonumber
  \times &&   \frac{2-\sqrt{2}}{4 \pi l_B^2}
            \exp \left[ - \frac{2-\sqrt{2}}{8 l_B^2}
            \left( {\bf R}+{\bf r}_h \right)^2  \right] \, .
\end{eqnarray}
This shows that the distribution for the relative coordinate
${\bf R}-{\bf r}_h$ is squeezed {\em at the expense\/} of that for
the coordinate ${\bf R}+{\bf r}_h$, and the variances are
$ \langle \tilde{0}| ({\bf R} \pm {\bf r}_h)^2 | \tilde{0} \rangle  =
   4(2 \pm \sqrt{2}) \, l_B^2 $.
The squeezing enhances the $e$--$h$ attraction which
will be used below for constructing a trial wave function
of the 2D magnetoexciton $X^-$.

Let us now perform the second unitary transformation corresponding
to the diagonalization of the operator \cite{remark}
$\hat{\pi}_{+} = A_e^{\dag}({\bf r}_1) + A_e^{\dag}({\bf r}_2) -
A_h({\bf r}_h)$.
This introduces a new state
$|\bar{0} \rangle = \bar{S} \tilde{S} |0 \rangle =
\bar{S} |\tilde{0} \rangle$,
which corresponds to the simultaneous diagonalization
of the operators $\hat{k}_{-}$ and $\hat{\pi}_{+}$;
the unitary operator $\bar{S}  =  \exp ( \Theta \bar{\cal L} )$,
where the generator $\bar{\cal L}   =
A^{\dag}_{h}({\bf r}_h) A^{\dag}_{e}({\bf R}) - A_e({\bf R}) A_h({\bf r}_h)$.
The transformations effectively introduce new coordinates,
$\{ {\bf r},{\bf R},{\bf r}_h \} \rightarrow \{ {\bf r},\ro_1,\ro_2 \}$,
where
$\ro_1 = \sqrt{2} \, {\bf R} - {\bf r}_h$
and
$\ro_2 = \sqrt{2} \, {\bf r}_h - {\bf R}$, which
can be presented in the matrix form
\begin{equation}
        \label{R-rho}
             \left( \begin{array}{c} \ro_1   \\
                                     \ro_2   \end{array}  \right)
  = \hat{F} \left( \begin{array}{c} {\bf R}  \\
                                    {\bf r}_h  \end{array} \right)
                  \quad , \quad
      \hat{F}= \left( \begin{array}{rr}
            \cosh  \, \Theta & - \sinh  \, \Theta   \\
          - \sinh  \, \Theta &   \cosh  \, \Theta
                       \end{array} \right)  \,
\end{equation}
with $\cosh  \Theta=\sqrt{2}$, $\sinh \Theta=1$;
the matrix $\hat{F}$ corresponds to the $SU(1,1)$ symmetry. \cite{Klauder}
Indeed, the inter-LL ladder operators are changed under the Bogoliubov
transformations as
\begin{equation}
        \label{A-Arho}
 \bar{S} \left( \begin{array}{c}  A^{\dag}_{e}({\bf R}) \\
                         A_{h}({\bf r}_h)  \end{array}  \right) \bar{S}^{\dag}
 = \hat{F} \left( \begin{array}{c} A^{\dag}_{e}({\bf R}) \\
                                   A_{h}({\bf r}_h)  \end{array} \right)
 =  \left( \begin{array}{c} A^{\dag}_{e}(\ro_1) \\
                                   A_{h}(\ro_2)  \end{array} \right) \, .
\end{equation}
The intra-LL operators (\ref{Bog1}) and (\ref{Bog2}) transform
according to the same representation.
The coordinate representation
\begin{equation}
   \label{vac-ro3}
   \langle {\bf r}\ro_1\ro_2| \bar{0} \rangle =
     \frac{1}{(2\pi l_B^2)^{3/2}}
   \exp \left( -\frac{{\bf r}^2+\ro_1^2+\ro_2^2}{4 l_B^2} \right)
\end{equation}
shows that $| \bar{0} \rangle$ is a {\em true vacuum\/} for both the
intra-LL $B^{\dag}_{e}(\ro_1)$, $B^{\dag}_{h}(\ro_2)$
and inter-LL  $A^{\dag}_{h}(\ro_2)$, $A^{\dag}_{e}(\ro_1)$ operators.
Now we can perform the change of the variables
$\{ {\bf r},{\bf R},{\bf r}_h \} \rightarrow \{ {\bf r},\ro_1,\ro_2 \}$
in the basis states:
\begin{eqnarray}
   \nonumber
 & & | n_{r} n_{R} n_{h}; \widetilde{ k l m} \rangle  = \\
   \nonumber
 & & \frac{ \bar{S}^{\dag}
                   A_e^{\dag}({\bf r})^{n_r}
                   A_e^{\dag}(\ro_1)^{n_R} A_h^{\dag}(\ro_2)^{n_h}
                   B_e^{\dag}(\ro_1)^{k}
                   B_h^{\dag}(\ro_2)^{l}
                   B_e^{\dag}({\bf r})^{m}
   | \bar{0} \rangle}{(n_{r}!n_{R}!n_{h}!k!l!m!)^{1/2}} \\
         \label{bas_new}
 & & \equiv  \bar{S}^{\dag} | \overline{ n_{r} n_{R} n_{h}; k l m} \rangle \, .
\end{eqnarray}
The overline shows that a state is generated in the usual way by the
intra- and inter-LL Bose ladder operators acting on the true vacuum
$|\bar{0}\rangle$ --- all in the representation of
the coordinates $\{{\bf r},\ro_1,\ro_2\}$.
The Hamiltonian $H$ is block-diagonal in the quantum numbers
$k,M_z$ (and $S_e$, $S_h$). Due to the Landau degeneracy \cite{Simon,Dz&S_PRL}
in $k$, it is sufficient to consider the states with $k=0$.
This effectively removes one degree of freedom
and corresponds to a partial separation of the CM motion.
From now on we will consider the $k=0$ states only,
designating such states in (\ref{bas_new}) as
$| \overline{ n_{r} n_{R} n_{h}; l m} \rangle$.
For the Hamiltonian we arrive therefore at the unitary
transformation
\begin{eqnarray}
       \nonumber
 &&\langle \widetilde{ m_2 l_2 }; n_{h2} n_{R2} n_{r2} | H
  | n_{r1} n_{R1} n_{h1}; \widetilde{ l_1 m_1 } \rangle      \\
  \label{me1}
  & &    = \langle \overline{ m_2 l_2 ; n_{h2} n_{R2} n_{r2} } |
               \bar{S} H \bar{S}^{\dag}
  | \overline{n_{r1} n_{R1} n_{h1}; l_1 m_1 } \rangle  \, ,
\end{eqnarray}
which is the main formal result of this work.

The Coulomb interparticle interactions (\ref{H_int})
in the coordinates \{${\bf r},\ro_1,\ro_2$\} take the form
\begin{equation}
        \label{Ham_rho}
H_{\rm ee} = \frac{e^2}{\sqrt{2} \epsilon r} \quad ,  \quad
H_{\rm eh} = - \frac{\sqrt{2}e^2}{\epsilon|\ro_2 - {\bf r}|}
            - \frac{\sqrt{2}e^2}{\epsilon|\ro_2 + {\bf r}|} \, ,
\end{equation}
and $H_{\rm int}$ {\em does not depend\/} on $\ro_1$.
From Eq.~(\ref{A-Arho}) it follows that the free Hamiltonians transform as
$\bar{S}  H_{0e}({\bf r})   \bar{S}^{\dag} = H_{0e}({\bf r})$,
$\bar{S}  H_{0e}({\bf R})   \bar{S}^{\dag} = H_{0e}(\ro_1)$,
and
$\bar{S}  H_{0h}({\bf r}_h) \bar{S}^{\dag} = H_{0h}(\ro_2)$
and describe {\em new effective particles} --- free $e$ and $h$ in a
magnetic field --- with the {\em modified interactions\/} (\ref{Ham_rho}).
The Hamiltonian of the $e$--$e$ interactions
$H_{\rm ee}({\sqrt{2}|{\bf r}|})$
does not depend on $\ro_1$, $\ro_2$ and, therefore, is invariant:
$\bar{S} H_{\rm ee} \bar{S}^{\dag} = H_{\rm ee}$.
Thus, the matrix elements of the $e$--$e$ interaction
are easily obtained from (\ref{me1}): they reduce to
the matrix elements $V_{n_1 m_1}^{n'_1 m'_1}$
describing the interaction of the electron
with a fixed negative charge $-e$:
\begin{eqnarray}
        \label{mat_ee}
 & & \langle \widetilde{ m_2 l_2 }; n_{h2} n_{R2} n_{r2} | H_{\rm ee}
  | n_{r1} n_{R1} n_{h1}; \widetilde{ l_1 m_1 } \rangle \\
     \nonumber
  & & \quad =  \langle \overline{ m_2 l_2 ; n_{h2} n_{R2} n_{r2}} | H_{\rm ee}
    | \overline{ n_{r1} n_{R1} n_{h1}; l_1 m_1 } \rangle   \\
     \nonumber
  & & \quad =  \delta_{n_{R1},n_{R2}}  \delta_{n_{h1},n_{h2}} \delta_{l_1,l_2}
 \delta_{n_{r1}-m_1,n_{r2}-m_2} \,
    \case{1}{\sqrt{2}} V_{n_{r1} \, m_1}^{n_{r2} \, m_2} \, .
\end{eqnarray}
In, e.g., zero \cite{SSC} LL
$V_{0 \, m}^{0 \, m}= [(2m-1)!!/2^m m!] \, E_0$,
where $E_0 = \sqrt{\frac{\pi}{2}} \, \frac{e^2}{\epsilon l_B}$.
The generator $\bar{\cal L}$ and the Hamiltonian $H_{\rm eh}({\bf r},\ro_2)$
do not form a closed algebra of a finite order.
Therefore, the explicit form of $\bar{S} H_{\rm eh} \bar{S}^{\dag}$
cannot be found. We can find, however, the form of the matrix elements
of $\bar{S} H_{\rm eh} \bar{S}^{\dag}$ in (\ref{me1}).
Because of the electron permutational symmetry
${\bf r} \leftrightarrow - {\bf r}$
it is sufficient to consider the term
$U_{\rm eh}(\ro_2-{\bf r}) = - e^2/\epsilon|\ro_2-{\bf r}|$.
Here we only consider the states in zero LL
$| \overline{000; l m} \rangle  \equiv |\overline{l m } \rangle$.
Using the normal-ordered form of $\bar{S}$, we have
\begin{eqnarray}
    \label{eh00}
& & \quad \langle \overline{m_2 l_2}| \bar{S} U_{\rm eh} \bar{S}^{\dag}
        |\overline{l_1 m_1} \rangle
  \equiv \bar{U}_{0m_1\,0l_1}^{0m_2\,0l_2} \\
       \nonumber
 & & = \case{1}{2} \langle \overline{m_2 l_2}|
  e^{-\case{1}{\sqrt{2}}A_e(\ro_1)A_h(\ro_2)}
           U_{\rm eh}
  e^{-\case{1}{\sqrt{2}}A_h^{\dag}(\ro_2)A_e^{\dag}(\ro_1)}
  |\overline{l_1 m_1} \rangle  \, .
\end{eqnarray}
Expanding the exponents and exploiting the fact that
$U_{\rm eh}(\ro_2-{\bf r})$ does not depend on $\ro_1$, we obtain a series
\begin{equation}
        \label{ser1} 
 \bar{U}_{0m_1\,0l_1}^{0m_2\,0l_2} =
      \case{1}{2} \sum_{p=0}^{\infty}
        \left( \case{1}{2} \right)^p U_{0m_1\,pl_1}^{0m_2\,pl_2} \, .
\end{equation}
Note that (\ref{ser1}) includes contributions of the
{\em infinitely many\/} LL's.
For the Coulomb interactions the matrix elements can be
calculated analytically; in zero LL we obtain
\begin{equation}
      \label{new-repr}
  \langle \overline{m_2 \, l_2}| \bar{S} H_{\rm eh} \bar{S}^{\dag}
                     | \overline{l_1 \, m_1}\rangle
  = \delta_{l_1-m_1,l_2-m_2} \, 2\sqrt{2} \,
   \bar{U}_{{\rm min}(m_1,m_2),{\rm min}(l_1,l_2)}(|m_1-m_2|) \, ,
\end{equation}
\begin{eqnarray}
        \label{barU1}
  & & \bar{U}_{mn}(s)  =
 - \frac{E_0}{\left[m!(m+s)!n!(n+s)!\right]^{1/2} \, 2^{m+n+s}3^{s+1/2}} \\
   \nonumber
   & & \times  \sum_{k=0}^{m} \, \sum_{l=0}^{n} \,
    C_m^k \, C_n^l \,  \left(\case{2}{3}\right)^{k+l} \,
    [2(k+l+s)-1]!! \, [2(m-k) -1]!!                   \\
    \nonumber
   & & \times  \sum_{p=0}^{n-l} \, C_{k}^{p} \, C_{n-l}^{p} \,
              (-1)^p \, p! \, [2(n-l-p)-1]!! \, .
\end{eqnarray}

The developed formalism can be used for performing
a rapidly convergent \cite{SSC} expansion of the
interacting $e$--$h$ states in the basis (\ref{me1}),
which preserves all symmetries of the problem.
Here we demonstrate a possibility of using the squeezed
states for constructing {\em trial\/} wave functions.
We consider the triplet charged 2D magnetoexciton
in zero LL, $X^-_{t00}$, with $M_z=-1$, which
is the only bound state \cite{X-th,Dz&S_PRL} in zero LL
in the strictly-2D system in the high-field limit.
The {\em simplest possible\/} wave function in zero LL
compatible with {\em all\/} symmetries of the problem is
\begin{equation}
   \label{zeroLL}
 \langle {\bf r}{\bf R}{\bf r}_h | B^{\dag}_{e}({\bf r}) |\tilde{0} \rangle =
          \frac{1}{\sqrt{2}\,(2\pi l_B^2)^{3/2}}
           \left( \frac{z^{\ast}}{\sqrt{2}l_B} \right)
      \exp \left( -\frac{{\bf r}^2+{\bf R}^2+{\bf r}_h^2-
      \sqrt{2} Z^{\ast}z_h }{4 l_B^2} \right) \, .
\end{equation}
This form allows analytic calculations and, as a squeezed state (see above),
already ensures the $X^-_{t00}$ binding.
Indeed, the total Coulomb interaction energy is given by
$\frac{1}{\sqrt{2}}V_{0 \, 1}^{0 \, 1} + 2\sqrt{2}\,\bar{U}_{1\,0}(0)=
(\frac{\sqrt{2}}{4}-\frac{5\sqrt{6}}{9}) E_0 \simeq -1.007E_0$.
The corresponding binding energy (counted from the ground state
energy of the neutral magnetoexciton, $-E_0$)
is $0.007E_0$, which is $17\%$ of the numerically
``exact'' value of $0.043E_0$. \cite{X-th,SSC}
A similar type of squeezing can be applied to construct
a trial wave function of the $X^{-}_{t00}$.
The idea is to additionally squeeze the effective
hole $\ro_2$ and electron ${\bf r}$ coordinates.
Since the wave function must by antisymmetric under the permutation
of the electron coordinates, we can use the form
$ | \phi \rangle \sim
B^{\dag}_e({\bf r})(S_{\phi} + S_{-\phi}) \bar{S}^{\dag}|\bar{0}\rangle$,
where the second two-mode squeezing operator is given by
$S_{\phi} =\exp[ \phi B^{\dag}_e({\bf r})B^{\dag}_h(\ro_2) - {\rm H.c.}]$
and we have used $|\tilde{0} \rangle = \bar{S}^{\dag} |\bar{0} \rangle$.
The normalized {\em four\/}-mode squeezed wave function has the form
\begin{equation}
         \label{probe}
 | \phi \rangle = \frac{1+\tanh^2\phi}{\cosh^2 \phi \,
                        \sqrt{1+\tanh^4\phi}} \,
     B^{\dag}_e({\bf r}) \, \cosh \left[ \tanh\phi \,
     B^{\dag}_e({\bf r})B^{\dag}_h(\ro_2)  \right]
     \bar{S}^{\dag} |\bar{0}\rangle \, .
\end{equation}
The calculated energy of the Coulomb $e$--$e$ repulsion,
$\langle \phi| H_{\rm ee}| \phi \rangle$, monotonically
decreases with increasing the transformation angle $\phi$,
whereas the energy of the $e$--$h$ attraction,
$-\langle \phi| H_{\rm eh}| \phi \rangle$, has a maximum (see Fig.~1).
The binding of the $X_{t00}^-$ results from a rather delicate
balance between the two terms, and for the state (\ref{probe})
the maximum achieved binding energy is $E_b \simeq 0.038 E_0$
(see inset to Fig.~1), which is 91\% of the ``exact'' value; \cite{SSC,X-th}
note that the inaccuracy is 0.3\% of the $e$--$h$
interaction energy. Similar type of squeezed trial
wave functions may be useful in other
solid state and atomic physics problems dealing with
correlated $e$--$h$ states in strong magnetic fields.

In conclusion, we have developed for charged $e$--$h$ systems in
magnetic fields an operator approach that allows one to partially
separate the CM motion. This results in the appearance of new effective
particles, electrons and holes in a magnetic field, with modified
interparticle interactions. A relation of the considered basis states
with the two-mode squeezed oscillator states has been established.

This work was supported in part by a COBASE grant
and by the Russian Ministry of Science program ``Nanostructures''.

\begin{figure}[!hb]
\caption{The expectation values
of the $e$--$e$ repulsion, $\langle \phi| H_{\rm ee} |\phi \rangle$,
and the $e$--$h$ attraction,
$\langle \phi| H_{\rm eh} |\phi \rangle$ (with the opposite sign,
counted from the neutral magnetoexciton binding energy
$E_0 = \protect\sqrt{\frac{\pi}{2}} \, \frac{e^2}{\epsilon l_B}$)
for the trial wave function (\protect\ref{probe})
of the charged triplet magnetoexciton in zero LL's, $X^-_{t00}$.
The binding energy
$E_b= - \langle \phi| H_{\rm eh}+H_{\rm ee} |\phi \rangle - E_0$
is shown in the inset.
}                \label{fig1}
\end{figure}


\begin{references}
\bibitem{remark}Note that the operators $\hat{\pi}_{\pm}$ do not commute
and, in general, do not form a simple algebra with the Hamiltonian.
A special case is when the charge-to-mass ratio
$e_j /m_j = \mathrm{const}$, and
$[H, \hat{\pi}_{\pm}] =  \mp \hbar (e_j B/m_jc) \, \hat{\pi}_{\pm}$,
which corresponds to the CM separation.
\end{references}
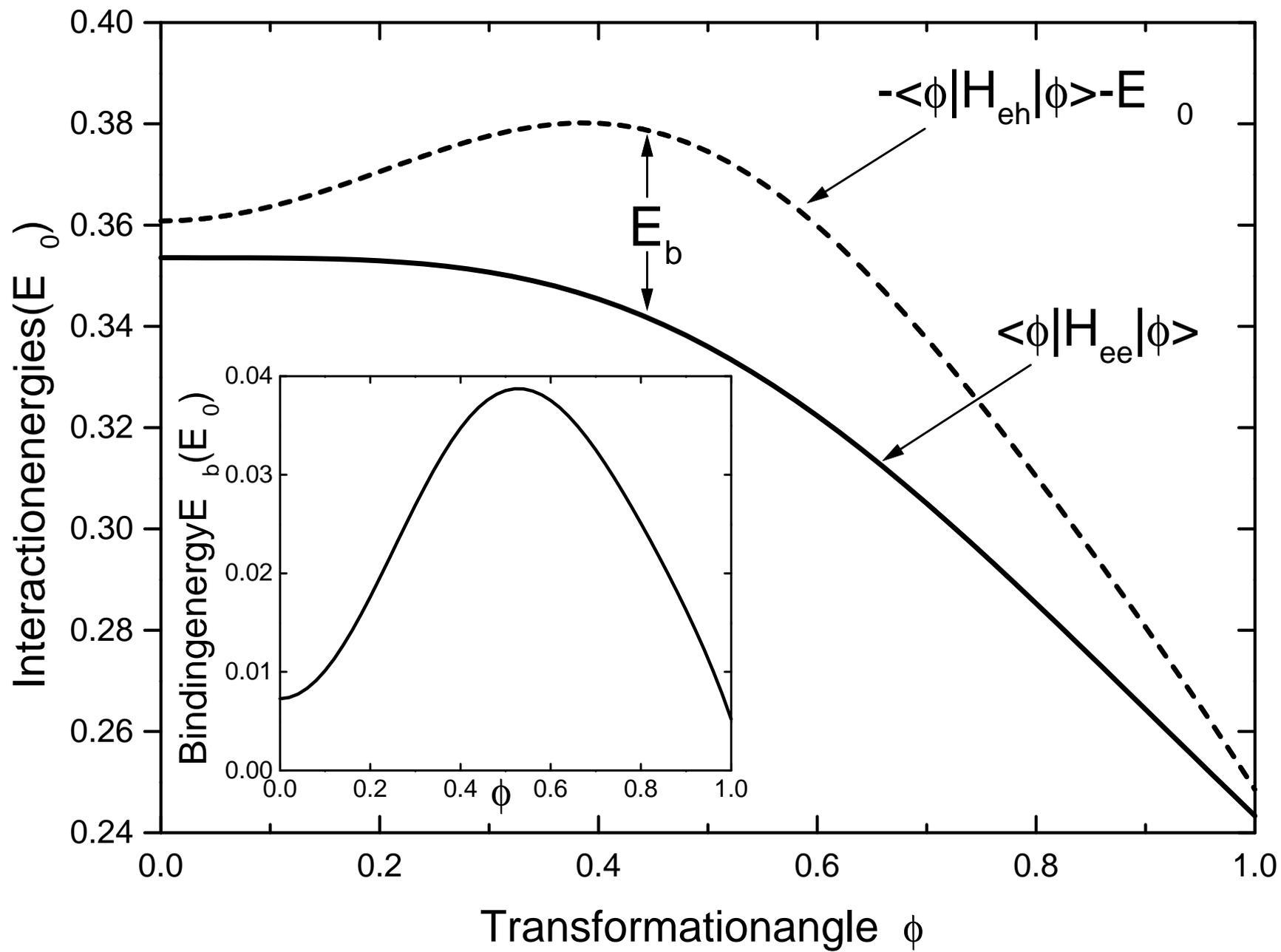
\end{document}